\documentclass{PoS}
\usepackage{array}
\def\dt{\delta\tau}		
\def\dH{\delta H}		
\def\trjlen{\tau}		
\def\defn{\equiv}
\def\hats{\hat S}		
\def\hatt{\hat T}		
\def\SST{\{S,\{S,T\}\}}
\def\TST{\{T,\{S,T\}\}}
\def\TTTST{\{T,\{T,\{T,\{S,T\}\}\}\}}
\def\STSST{\{\{S,T\},\{S,\{S,T\}\}\}}
\def\TSSST{\{T,\{S,\{S,\{S,T\}\}\}\}}
\def\TTSST{\{T,\{T,\{S,\{S,T\}\}\}\}}
\def\SSSST{\{S,\{S,\{S,\{S,T\}\}\}\}}
\def\STTST{\{\{S,T\},\{T,\{S,T\}\}\}}
\def\O{{\cal O}}		
\def\Tr{\mathop{\rm Tr}}	
\def\Re{\mathop{\rm Re}}	

\def\rational#1#2{{\mathchoice{\textstyle{#1\over#2}}%
  {\scriptstyle{#1\over#2}}{\scriptscriptstyle{#1\over#2}}{#1/#2}}}
\def\half{\rational12}		
\def\quarter{\rational14}	

\def\dd#1#2{{\mathchoice{d#1\over d#2}%
  {d#1\over d#2}%
  {d#1\!/\!d#2}%
  {d#1\!/\!d#2}}}		
\def\pdd#1#2{{\mathchoice{\partial#1\over\partial#2}%
  {\partial#1\over\partial#2}%
  {\partial#1\!/\!\partial#2}%
  {\partial#1\!/\!\partial#2}}}	
\def\ddq{\pdd{}q}		
\def\ddp{\pdd{}p}		
\title{Speeding up HMC with better integrators}
\ShortTitle{Speeding up HMC}
\author{
  \speaker{A. D. Kennedy}\\
    School of Physics, The University of Edinburgh \& SUPA,\\
    Mayfield Road, Edinburgh, EH9 3JZ, United Kingdom\\
    E-mail: \email{adk@ph.ed.ac.uk}}
\author{M. A. Clark\\
    Center for Computational Sciences, Boston University,\\
    3 Cummington Street, Boston, MA 02215, United States of America\\
    E-mail: \email{mikec@bu.edu}}
\abstract{We discuss how dynamical fermion computations may be made yet cheaper
by using symplectic integrators that conserve energy much more accurately
without decreasing the integration step size.  We first explain why symplectic
integrators exactly conserve a ``shadow'' Hamiltonian close to the desired one,
and how this Hamiltonian may be computed in terms of Poisson brackets.  We then
discuss how classical mechanics may be implemented on Lie groups and derive the
form of the Poisson brackets and force terms for some interesting integrators
such as those making use of second derivatives of the action (Hessian or force
gradient integrators).  We hope that these will be seen to greatly improve
energy conservation for only a small additional cost and that their use will
significantly reduce the cost of dynamical fermion computations.}
\FullConference{The XXV International Symposium on Lattice Field Theory\\
		 July 30-4 August 2007\\
		 Regensburg, Germany}
\begin{document}

\section{Symplectic Integrators}

We are interested in finding the classical trajectory in phase space of a
system described by the Hamiltonian \(H(q,p) = T(p) + S(q) = \half p^2 +
S(q)\). The idea of a \emph{symplectic integrator} is to write the time
evolution operator as \[\exp\left(\tau\dd {}t\right) = \exp\left(\tau\left\{
\dd pt \ddp + \dd qt\ddq\right\}\right) \defn e^{\tau\hat H}\] where the
\emph{vector field} \[\hat H = -\pdd Hq\ddp + \pdd Hp\ddq = -S'(q)\ddp +
T'(p)\ddq \defn \hats + \hatt.\] Since the kinetic energy \(T\) is a function
only of \(p\) and the potential energy \(S\) is a function only of \(q\) it
follows that the action of \(e^{\tau\hats}: f(q,p) \mapsto f(q,p - \tau
S'(q))\) and \(e^{\tau\hatt}: f(q,p) \mapsto f(q + \tau T'(p),p)\) are just
translations of the appropriate variable.

We now make use of the Baker--Campbell--Hausdorff (BCH) formula, which tells us
that the product of exponentials in any associative algebra can be written as
\(\ln(e^{A/2}e^Be^{A/2}) - (A + B) = \rational1{24}\left\{\strut [A,[A,B]] -
2[B,[A,B]]\right\} +\cdots\) where all the terms on the right hand side are
constructed out of commutators of \(A\) and \(B\) with known coefficients.  We
find that for a simple PQP symmetric integrator with step size \(\dt\) the
evolution operator for a trajectory of length \(\trjlen\) may be written as
\begin{eqnarray*}
  U_{\mbox{\tiny PQP}}(\dt)^{\trjlen/\dt} 
  &=& \left(e^{\half\dt\hats} e^{\dt\hatt}
    e^{\half\dt\hats}\right)^{\trjlen/\dt} \\
  &=& \left(\exp\left[(\hatt + \hats)\dt 
    - \rational1{24}\left(\strut[\hats, [\hats,\hatt]] 
    + 2[\hatt,[\hats,\hatt]]\right) \dt^3
    + \O(\dt^5)\right]\right)^{\trjlen/\dt} \\
  &=& \exp\left[\trjlen\left(\hatt + \hats
    - \rational1{24}\left(\strut[\hats, [\hats,\hatt]]
    + 2[\hatt, [\hats,\hatt]]\right) \dt^2
    + \O(\dt^4)\right)\right].
\end{eqnarray*}

\section{Shadow Hamiltonians} \label{sec:shadow}

For every symplectic integrator there is a \emph{shadow Hamiltonian} \(\tilde
H\) that is exactly conserved; this may be obtained by replacing the
commutators \([\hats,\hatt]\) in the BCH expansion with the \emph{Poisson
bracket} \(\{S,T\} \defn \pdd Sp\pdd Tq - \pdd Sq\pdd
Tp\)~\cite{inexact:2007}. For example our PQP integrator above exactly
conserves the shadow Hamiltonian \(\tilde H \defn T + S -
\rational1{24}\left(\strut\SST + 2\TST \right)\dt^2 + \cdots\).

We now make the simple observation that any symplectic integrator is
constructed from the same Poisson brackets, and that these Poisson brackets are
extensive quantities. We therefore propose to measure the average values of the
Poisson brackets and then optimize the integrator (by adjusting the step sizes,
order of the integration scheme, integrator parameters, number of pseudofermion
fields, etc.~\cite{Clark:2006fx,forcrand:2006}) offline so as to minimize the
cost. This is possible because the acceptance rate and instabilities are
completely determined by \(\dH = \tilde H - H\).

As a very simple example consider the minimum norm PQPQP integrator
\[U_{\mbox{\tiny{PQPQP}}}(\dt)^{\tau/dt} = \left(e^{\alpha\hats\dt}
e^{\half\hatt\dt} e^{(1-2\alpha)\hats\dt} e^{\half\hatt\dt}
e^{\alpha\hats\dt}\right)^{\tau/dt}\] whose shadow Hamiltonian is \[\tilde H =
H + \left(\frac{6\alpha^2 - 6\alpha + 1}{12} \SST + \frac{1 - 6\alpha}{24} \TST
\right)\dt^2 + \O(\dt^4).\] With only one degree of freedom \(\alpha\) we
cannot completely eliminate the coefficient of the \(O(\dt^2)\) contribution,
however, we may optimize this integrator by setting the parameter \(\alpha =
\half + \quarter \frac{\left\langle \TST \right\rangle} {\left\langle \SST
\right\rangle}\). There have been alternative optimization strategies proposed:
minimizing the \(L_2\) norm of coefficients assuming \(|\SST| = |\TST|\)
\cite{omelyan:2003}, and setting the coefficient of one of the two Poisson
brackets to zero by choosing \(\alpha=\frac{1}{2}(1-\frac{1}{\sqrt{3}})\) or
\(\frac{1}{6}\).  However, these strategies clearly break down when optimizing
higher order minimum norm integrators, i.e., for \(O(\dt^4)\) integrators there
are 6 Poisson bracket contributions that must be considered (see
Table~\ref{table:3}).

\section{Hessian Integrators}

We now make another simple observation: consider again the PQPQP integrator,
where we set \(\alpha=\frac{1}{6}\) so that the \(\TST\) contribution is
eliminated. The remaining leading order Poisson bracket \(\SST\) depends only
on \(q\), which means that we can evaluate the integrator step
\(e^{\widehat{\SST}\dt^3}\) explicitly (it is again just a shift of \(p\)). The
force for this integrator step involves second derivatives of the action, and
therefore they are called Hessian or force gradient integrators
\cite{chin:2000,omelyan:2002}. By putting such an integration step into a
multistep integrator we can eliminate all the leading \(\O(\dt^2)\) terms in
\(\dH\). The advantage of such an integrator over that of Campostini
\cite{campostrini89a, creutz89a} is that the coefficients of the next order
terms are approximately two orders of magnitude smaller (see
Table~\ref{table:3}). We want to stress that although eliminating the leading
term must be best asymptotically as \(\dt\to0\) it might well not be the
optimal solution in practice; the optimal solution may be obtained by
minimizing \(\dH\) as discussed in \S\ref{sec:shadow}.

\section{Beyond Scalar Field Theory}

We now have to construct the Poisson brackets and Hessian integrators for gauge
fields, where the field variables are constrained to live on a group
manifold. To do this we need to use some differential geometry.
Table~\ref{table:1} summarizes the difference between the formulation on flat
space that we have discussed up to this point and that on general manifolds.

\begin{table}
  \begin{center}
    \begin{tabular}{c|c|c}
      & Flat Manifold & General \\
      \hline
      Symplectic 2-form & \(dp\wedge dq\) & \(\omega : d\omega = 0\)\\
      Hamiltonian vector field & \(\hat H = \pdd Hp\ddq - \pdd Hq\ddp\) &
        \(dH = i_{\hat H}\omega\) \\
      Equations of motion & \(\dot q = \pdd Hp, \dot p = -\pdd Hq\) &
        \(\left.\dd{}t\right|_\sigma = \hat H\)\\
      Poisson bracket & \(\{A,B\} = \pdd Ap\pdd Bq - \pdd Aq\pdd Bp\) &
        \(\{A,B\} = -\omega(\hat A,\hat B)\)
    \end{tabular}
  \end{center}
  \caption{Comparison of quantities in flat space and on a general 
    manifold~\cite{inexact:2007}.}
  \label{table:1}
\end{table}

In order to construct a Hamiltonian system on a manifold we need not only a
Hamiltonian function but also a fundamental closed 2-form \(\omega\).  On a Lie
group manifold this is most easily found using the globally defined
\emph{Maurer--Cartan} forms \(\{\theta^i\}\) that are dual to the generators
and satisfy the relation \(d\theta^i = -\half c^i_{jk} \theta^j\wedge
\theta^k\), where \(c^i_{jk}\) are the structure constants of the group.  We
choose to define \(\omega \defn -d\sum_i \theta^ip^i = \sum_i(\theta^i\wedge
dp^i - p^id\theta^i) = \sum_i (\theta^i \wedge dp^i + \half p^i c^i_{jk}
\theta^j \wedge\theta^k)\).  Using this fundamental 2-form we can define a
Hamiltonian vector field \(\hat A\) corresponding to any 0-form \(A\) through
the relation \(dA = i_{\hat A}\omega\), and in the natural coordinates \((e_i,
\dd{}{p^i})\) on the contangent bundle this gives
\begin{equation}
\hat A = \sum_i \left(\pdd A{p^i}e_i 
  + \left[\sum_{jk} c^k_{ji} p^k \pdd A{p^j}
    - e_i(A)\right] \pdd{}{p^i}\right).
  \label{eq:hamvec}
\end{equation}
The classical trajectories \(\sigma_t = (Q_t,P_t)\) are then the integral
curves of this vector field, \(\dot\sigma_t = \hat A(\sigma_t)\).

\section{Putting It All Together}

Recalling that \(H = S + T\) we can compute the Hamiltonian vector fields
corresponding to \(S\) and \(T\) using equation~(\ref{eq:hamvec}), and from
these we can evaluate the lowest-order Poisson bracket \[\{S,T\} = -
\omega(\hats, \hatt) = -(\theta^i\wedge dp^i + \half p^ic^i_{jk}\theta^j\wedge
\theta^k)(\hats,\hatt) = -p^ie_i(S) = -\Re\Tr\left(\pdd SUPU\right),\] and the
Hamiltonian vector field corresponding to it,
\begin{eqnarray*}
\widehat{\{S,T\}} 
  &=& \sum_i\left(\pdd{\{S,T\}}{p^i} e_i 
    + \left[\sum_{jk} c^k_{ji} p^k \pdd{\{S,T\}}{p^j} - e_i(\{S,T\})\right]
      \pdd{}{p^i}\right)\\
  &=& -e_i(S)e_i + \left[-c^k_{ji} p^k e_j(S) + p^je_ie_j(S)\right]\pdd{}{p^i}.
\end{eqnarray*}
From this we can derive expressions for the third- and fifth-order Poisson
brackets that are needed for symmetric symplectic integrators, and these are
listed in Table~\ref{table:2}. Similarly, we can then evaluate the
corresponding Hamiltonian vector fields for any Poisson brackets we wish to
include in the integration (e.g., \(\widehat{\SST}\) for force gradient
integrators).

\begin{table}
  \begin{center}
    \begin{tabular}{c|c}
      \(\SST\) & \(e_i(S)e_i(S)\) \\
      \(\TST\) & \(-p^ip^ke_ie_j(S)\) \\
      \(\SSSST\) & \(0\) \\
      \(\STSST\) & \(-2e_i(S)e_j(S)e_ie_j(S)\) \\
      \(\STTST\) & 
        \(\begin{array}{l}
	  3c^i_{jk}p^ip^\ell e_j(S) [e_ke_\ell(S) + e_\ell e_k(S)] \\
          \qquad + p^ip^j\left[\strut e_k(S)e_ke_ie_j(S)
	    - [e_ke_i(S) + e_ie_k(S)]e_ke_j(S)\right]
	\end{array}\) \\
      \(\TSSST\) & \(0\) \\
      \(\TTSST\) & \(2p^ip^j[e_ie_je_k(S)e_k(S) + e_ie_k(S)e_je_k(S)]\) \\
      \(\TTTST\) & \(-p^ip^jp^kp^\ell e_ie_je_ke_\ell(S)\) \\
    \end{tabular}
  \end{center}
  \caption{Poisson brackets required for symmetric symplectic integrators.}
  \label{table:2}
\end{table}

The explicit form of the shadow Hamiltonian for a variety of integrators is
show in Table~\ref{table:3}.

\begin{table}
  \begin{center}
    \setbox0=\hbox{} \ht0=0pt \dp0=0pt
    \newif\ifexp \expfalse
    \def\1#1#2{\ifexp\exp\left(#1\dt\,\hat#2\right)\else e^{#1\dt\,\hat#2}\fi}
    \def\2#1#2{\(#1\;#2\;#1\)}
    \def\3#1#2#3{\(\begin{array}{c}
          #1\;#2 \\
          \times\;#3 \\
	  \times\;#2\;#1
	\end{array}\)}
    \def\4#1#2#3#4{\(\setlength{\extrarowheight}{2ex}\exptrue\begin{array}{c}
	  #1\\
	  \times\;#2\\
	  \times\;#3 \\
          \times\;#4 \\
	  \times\;#3 \\
	  \times\;#2 \\
	  \times\;#1
	\end{array}\)\expfalse}
    \def\7#1#2#3#4{\(\setlength{\extrarowheight}{2ex}\begin{array}{c}
	  #1\;#2\;#3 \\
          \times\;#4 \\
	  \times\;#3\;#2\;#1
	\end{array}\)}
    \def\5#1#2#3{\mathchoice{{\textstyle\6{#1}{#2}{#3}}}%
      {{\scriptstyle\6{#1}{#2}{#3}}}%
      {{\scriptscriptstyle\6{#1}{#2}{#3}}}%
      {{\scriptscriptstyle\6{#1}{#2}{#3}}}}
    \newif\ifprev
    \newif\ifunit
    \def\8#1#2{
      \ifcat x#2x\unittrue\else\unitfalse\fi
      \count0=#1
      \ifnum\count0=0\else
        \ifnum\count0>0 \ifprev+\fi
	  \ifunit\the\count0\else\ifnum\count0>1 \the\count0\fi\fi
	\fi
	\count0=-\count0
        \ifnum\count0>0 -
	  \ifunit\the\count0\else\ifnum\count0>1 \the\count0\fi\fi
	\fi
	\ifunit\else#2\fi\prevtrue
      \fi}
    \def\6#1#2#3{\8{#1}{\root3\of4}\8{#2}{\root3\of2}\8{#3}{}}
    \begin{tabular}{c|c|c}
      Integrator & Update steps & Shadow Hamiltonian \\
      \noalign{\hrule height1.5pt}
      PQP & \2{\1\half S}{\1{}T}
        & \(T + S - \frac{\dt^2}{24} \left(\strut\SST + 2\TST\right)\) \\
      \hline
      QPQ & \2{\1\half T}{\1{}S}
        & \(T + S + \frac{\dt^2}{24} \left(\strut 2\SST + \TST\right)\) \\
      \hline
      \begin{tabular}{c}
	PQPQP \\
	\(\alpha=\frac{1}{6}\) \\
	\cite{omelyan:2002,omelyan:2003,forcrand:2006}
      \end{tabular}
      & \3{\1{\rational16}S}{\1\half T}{\1{\rational23}S}
      & \(T + S + \frac{\dt^2}{72}\SST\) \\
      \hline
      \begin{tabular}{c}
	PQPQP \\
	\(\alpha=\frac{1}{2}(1-\frac{1}{\sqrt{3}})\) \\
	\cite{omelyan:2002,omelyan:2003,forcrand:2006}
      \end{tabular}
      & \3{\1{\frac{3-\sqrt3}6}S}{\1\half T}{\1{\frac1{\sqrt3}}S}
      & \(T + S + \frac{\sqrt3-2}{24}\dt^2\TST\) \\
      \hline
      \begin{tabular}{c}
	Campostrini \\
	\cite{campostrini89a,creutz89a}
      \end{tabular}
      & \4{\1{\frac{\5{1}{2}{4}}{12}}T}%
          {\1{\frac{\5{1}{2}{4}}6}S}%
          {\1{\frac{\5{-1}{-2}{2}}{12}}T}%
          {\1{-\frac{\5{1}{2}{1}}3}S}
      & \(\begin{array}{c}
	  T + S \\
	  + \rational{\dt^4}{34560} \left(
	  \begin{array}{c}
	    -(\5{40}{40}{48})\;\SSSST\\
	    +(\5{180}{240}{312})\;\STSST \\
	    +(\5{60}{80}{104})\;\STTST\\
	    +(\5{-20}{0}{8})\;\TSSST \\
	    +(\5{0}{20}{32})\;\TTSST \\
	    +(\5{0}{5}{8})\;\TTTST
	  \end{array}\right)
        \end{array}\) \\
      \hline
      \begin{tabular}{c}
	Force \\ Gradient \\ \#1 \\ \cite{chin:2000,omelyan:2002}
      \end{tabular}
      & \7{\1{\rational16}T}%
	  {\1{\rational38}S}%
	  {\1{\rational13}T}%
	  {e^{\frac{48\dt\;S -\dt^3\;\widehat{\SST}}{192}}}
      & \(\begin{array}{c}
	  T + S \\
	  + \rational{\dt^4}{6635520} \left(
	  \begin{array}{c}
	    2259\;\SSSST \\
	    + 3024\;\STSST \\
	    + 768\;\STTST \\
	    + 5616\;\TSSST \\
	    + 4224\;\TTSST \\
	    + 896\;\TTTST
	  \end{array}\right)
        \end{array}\) \\
      \hline
      \begin{tabular}{c} Force \\ Gradient \\ \#2 \end{tabular}
      & \3{\1{\rational6}S}%
	  {\1{\rational2}T}%
	  {e^{\frac{48\dt\;S - \dt^3\;\widehat{\SST}}{72}}}
      & \(\begin{array}{c}
	  T + S \\
	  - \rational{\dt^4}{155520} \left(
	  \begin{array}{c}
	    41\;\SSSST \\
	    + 36\;\STSST \\
	    + 72\;\STTST \\
	    + 84\;\TSSST \\
	    + 126\;\TTSST \\
	    + 54\;\TTTST
	  \end{array}\right)
        \end{array}\) \\
      \noalign{\hrule height1.5pt}
    \end{tabular}
  \end{center}
  \vspace{-5mm}
  \caption{A collection of integrators with the leading terms in their
    exactly conserved shadow Hamiltonians.}
  \label{table:3}
\end{table}

\section{Conclusions}

Our work in this area is still very preliminary, so far we have concentrated on
developing these ideas. Future work shall focus on implementing and testing the
performance of these integrators for dynamical fermion calculations. We expect
that modest gains in performance can be expected through directly measuring the
leading order Poisson brackets to optimize the minimum norm family of
integrators. However, we hope that very significant performance improvements
can be obtained from force gradient integrators.

\acknowledgments
This work was supported in part by NSF grant PHY-0427646.

\iffalse
  \bibliography{adk,lattice-bibliography}
  \bibliographystyle{unsrt}
\else
  
\fi

\end{document}